# Observation and research on cosmic ray muons and solar modulation effect based on plastic scintillator detector[*]

WANG Dexin[1, 2], ZHANG Rui[1], YU Dekang[3], NA Hui[1], YAO Zhanghao[1], WU Linghe[1], ZHANG Suyalatu[1, 2, *], LIANG Tairan[1], HUANG Meirong[1, 2], WANG Zhilong[1, 2], BAI Yu[1, 2], HUANG Yongshun[1, 2], YANG Xue[1, 2], ZHANG Jiawen[3], LIU Mengdi[3], MA Qiang[3], YU Jing[3], JI Xiuyan[3], YU Yiliqi[3], SHAO Xuepeng[3, *]

1.College of Physics and Electronics, Inner Mongolia Minzu University, Tongliao 028043, China

2.Institute of Nuclear Physics, Inner Mongolia Minzu University, Tongliao 028043, China

3.Tongliao Meteorological Bureau, Tongliao 028000, China

**Abstract**

Cosmic rays, originating from stars, supernovae, and other astrophysical sources, are composed of high-energy particles that enter Earth's atmosphere. Upon interaction with atmospheric nuclei, these primary cosmic rays generate secondary particles, including neutrons, electrons, and muons, with muons constituting a dominant component at ground level. Muons, due to their relative abundance, stability, and well-characterized energy loss mechanisms, serve as critical probes for investigating the fundamental properties of cosmic rays. Studies of muon energy distribution, diurnal anisotropy, and their modulation by solar activity provide critical insights into the mechanism of particle acceleration in cosmic ray sources and the effects of solar and atmospheric.This study aims to characterize the counting spectra and anisotropic properties of cosmic ray muons by using a plastic scintillator detector system. The experiment was conducted over a three-month period, from December 2023 to February 2024, leveraging long-bar plastic scintillator detectors equipped with dual-end photomultiplier tubes (PMTs) and a high-resolution digital data acquisition system. A dual-end coincidence measurement technique was used to enhance the signal-to-noise ratio by suppressing thermal noise and other background interferences. Comprehensive

---

calibration of the detection system was performed using standard gamma-ray sources, including $^{137}$Cs, $^{60}$Co, and $^{40}$K, thereby ensuring precise energy scaling and reliable performance.The observed energy spectra of cosmic ray muons are in excellent agreement with theoretical predictions, which explains the energy losses caused by muons passing through the detector. Diurnal variations in muon count rates exhibit a pronounced pattern, with a systematic reduction occurring between 8:00 AM and 1:00 PM. This phenomenon is attributed to the solar shielding effects, where enhanced solar activity during daytime hours modulates the flux of galactic cosmic rays reaching Earth's surface. To account for atmospheric influences, meteorological corrections are performed using temperature and pressure adjustment functions derived from regression analysis. These corrections indicate that atmospheric pressure and temperature are significant factors affecting muon count rates, and a clear linear relationship is observed.The study further corroborates these findings through cross-comparisons with data from the Yangbajing Cosmic Ray Observatory. Minor discrepancies, primarily in low-energy muon count rates, are attributed to variations in detector sensitivities and local atmospheric conditions. These observations underscore the robustness of the plastic scintillator detector system for capturing detailed muon spectra and anisotropic patterns.This research establishes a reliable experimental framework for analyzing cosmic ray muons and their modulation by solar and atmospheric phenomena. The results contribute to a more in-depth understanding of anisotropy of cosmic rays and the interaction between astrophysical and geophysical processes. Furthermore, these findings provide valuable insights for optimizing detection technologies and enhancing the accuracy of cosmic ray studies.

# 1. Introduction

Cosmic ray is a kind of high-speed and high-energy particle stream, which mainly comes from stars, supernova remnants and active galactic nuclei outside the Milky Way[1]. When the primary cosmic ray enters the earth's atmosphere, it collides with atmospheric molecules to produce secondary particles, including muons, neutrons and electrons[2,3]. After interacting with environmental factors such as geomagnetic field and solar wind, these secondary particles show complex energy distribution, intensity variation and anisotropy characteristics[4,5]. The study of cosmic rays is not only helpful to reveal the mechanism of particle acceleration, but also of great significance to understand the physical processes of solar activity and terrestrial space environment[1,2].

In the study of cosmic ray secondary particles, muons are one of the main components in surface detection and have relatively stable characteristics, so they become an important object in the study of cosmic ray characteristics[6–11]. For example, Liu Jun and other[7,8] used wavelet transform and folding period analysis methods to analyze the meteorological effect of cosmic ray data from Yangbajing Observatory, and revealed the diurnal variation of Muzi flow intensity and its close relationship with temperature and pressure changes. Relevant studies have shown that meteorological factors have a significant impact on the surface muon count, especially the fluctuation of air pressure and temperature can cause the periodic change of the muon count. Therefore, in order to improve the reliability of observation data, these meteorological effects need to be corrected. In addition, the study also shows that solar activity affects the distribution and intensity of secondary particles through the modulation of cosmic rays, which provides an important clue to reveal the effect of solar modulation[6].

In recent years, plastic scintillators have been widely used in cosmic ray muon detection due to their low cost, high detection efficiency and excellent time response[12–16]. Compared with other types of detectors, plastic scintillator has the advantages of fast response time and stable signal, which is suitable for real-time measurement and large area detection[12]. In muon measurement, He Weijie and Li Bo developed a detector for muon lifetime measurement using plastic scintillator and photomultiplier tube[15], and achieved nanosecond time resolution. Wang Qiqi et al. Quantitatively analyzed the photon yield in the plastic scintillating fiber in the muon detection experiment[12], and proved that the photon counting method of scintillator can be used to measure the low-energy photons of muons.

However, the systematic application of plastic scintillators in the study of cosmic ray anisotropy and meteorological effects is still rare. In this paper, the plastic scintillator detector is used to observe the energy distribution and diurnal periodic variation of cosmic ray muons. Noise interference is effectively reduced by double-ended coincidence measurement and energy calibration, and the accuracy of measurement data is improved by combining with meteorological effect correction. The purpose of this study is to reveal the diurnal periodic variation of muon counts, to explore the manifestation of solar modulation effect in cosmic ray muons, and to provide experimental basis for understanding the anisotropy and modulation mechanism of cosmic rays.

## 2. Experiment and data processing

2.1 Experimental setup

In this experiment, the HND-S2 double-end output long strip plastic scintillator detector produced by Beijing Gaoneng Kedi Company is used. Its size is 1000 mm × 50 mm×50 mm. It is made of polystyrene, p-terphenyl and wave-shifting agent. The main peak of the emission spectrum is 423 nm. It has fast time response detection. The pulse rise time is 0.7 ns, the decay time is 2.6 ns, the pulse full width at half maximum is 1.8 ns, and the average hydrogen-carbon atomic ratio in the plastic scintillator is $1:1$[17]. The surface of the plastic scintillator is wrapped with a layer of aluminum foil with a thickness of 50 μm as a reflective layer to increase the reflectivity of the light on the inner surface of the plastic scintillator and improve the light collection efficiency. At the same time, a layer of black tape is wrapped on the outside to avoid the influence of natural light on the experimental measurement. Both ends of the scintillator are coupled with Hamamatsu CR105-02 end-window photomultiplier tubes (PMT), and the PMTs at both ends are powered by five-way negative high-voltage modules produced by Haiyang Bochuang Company. A desktop digital data acquisition DT5720B produced by CAEN, Italy, was used for data acquisition. The device has four channels, the maximum signal amplitude is $2V_{pp}$ ($V_{pp}$ is the amplitude of the oscilloscope waveform signal), and the sampling rate is up to 250 MS/s. With COMPASS data processing software, it can set the signal threshold, long/short gate, signal length and other parameters, and analyze the data online and offline. The schematic diagram of the experimental setup is shown in Fig. 1.

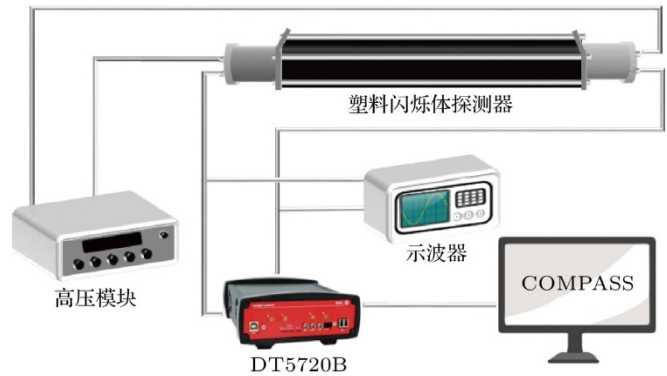

**Figure 1.** Schematic diagram of experimental setup

During the experiment, the adjustment of signal threshold and PMT voltage is very important for effective data acquisition. The signal waveform is observed by the oscilloscope. Considering the performance of the PMT and the parameters of the data acquisition system, the voltage of the PMT is finally set to – 785 V, and the signal threshold is set to 100 mV. The over-threshold signal of the PMTs at both ends on the oscilloscope is shown as Fig. 2.

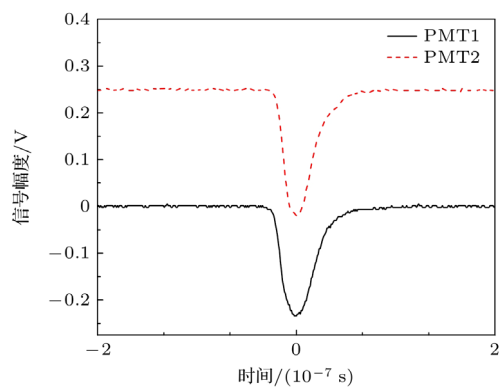

**Figure 2.** The over threshold signal of PMT at both ends output on the oscilloscope

After the input voltage is determined, the signal is directly input into the DT5720B, and the two-terminal PMT signal is directly analyzed and processed in the COMPASS software. Because of the thermal noise and afterwave effect on the PMT, and the coincidence characteristic of the cosmic ray muon signal at both ends of the scintillator in time, the double-ended readout mode is used in this experiment to reduce the noise interference of the single-ended readout detector. This double-ended readout method is often used in[17] in long strip plastic scintillator detectors. The coincidence time window of correlated events is studied by coincidence measurement technology. It is found that the appropriate coincidence time window setting can improve the detection efficiency of true events and reduce the influence of background noise. In addition, the particle track detector (PPAC) in the domestic large high energy physics experiment terminal RIBLL has also used a similar two-terminal readout technology[18–20] to improve the detection accuracy and reduce noise interference. These experiments effectively distinguish the real signal from the noise signal by setting readout devices at both ends of the detector and using the coincidence measurement principle, which is similar to the principle and purpose of the double-ended readout mode in this experiment.

Because the propagation time of the light emitted by the interaction between the muon and the material is about 5 ns in the 1 m long scintillator and the rise time of the PMT is 7 ns, considering the response of the data acquisition, it is finally determined that only when the two PMTs measure the signal at the same time within the time range of 15 ns, it is recognized as a valid muon event and recorded by the data acquisition. The signal was processed online using (digital pulse processing firmware) to obtain the muon counting spectrum.

2.2 Energy scale

In order to improve the accuracy of muon energy measurement,$^{137}$Cs and$^{60}$Co standard$\gamma$ sources were used to calibrate the energy of plastic scintillator detector. Because the energy resolution of the detector will lead to a certain broadening of the optical output spectrum, it is impossible to directly determine the maximum energy of Compton electrons, so the relationship between channel site and energy can be determined by using the position of the Compton edge in the optical output spectrum of the γ. The maximum energy of the Compton recoil electron Ec is calculated from the following equation,[21]:

$$E_c = E_\gamma \frac{\frac{2E_\gamma}{m_e c^2}}{1+\frac{2E_\gamma}{m_e c^2}}, \qquad (1)$$

Where Eγ is the incident energy of the standard radioactive source and $m_e c^2 = 0.511 \text{MeV}$ is the energy of the electron. In addition to the standard gamma sources $^{137}$Cs and $^{60}$Co, the measured optical output spectra were fitted by a Gaussian with $^{40}$K in the natural environmental background, and the results are shown in the Fig. 3(a). The exact position of the maximum Compton electron energy is determined at 81% of the maximum height of the Compton edge distribution[22]. The results show that the calibration points of the detector at 477, 1116 and 1243 keV are in good agreement with the theoretical values. Finally, the first order linear formula is used to fit the data, and the energy calibration curve is shown in Fig. 3(b).

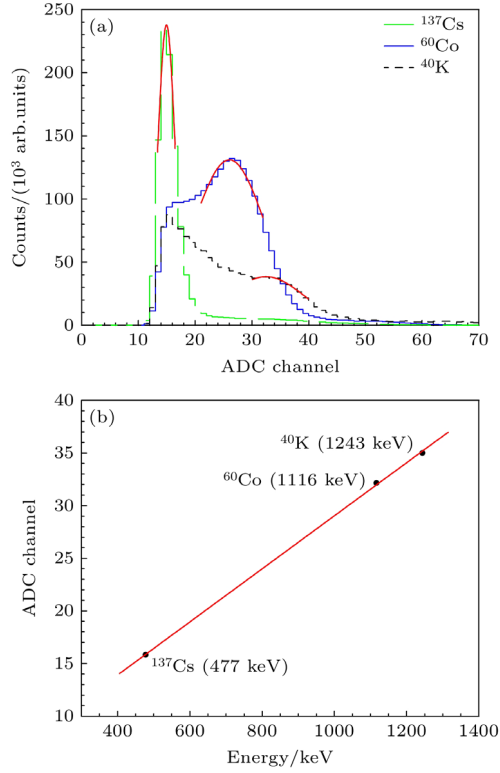

**Figure 3.** (a) Standard γ source light output spectrum; (b) energy scale curve

The measurement time of this experiment lasts from December 9, 2023 to February 11, 2024. Using the circular storage function of COMPASS software, the current counting spectrum data is automatically stored in txt and root file formats every hour. A total of 1424 files were collected in this experiment, and the root software package developed by CERN was used to process and analyze the experimental data offline.

2.3 Meteorological effect correction

When cosmic rays enter the earth's atmosphere, the conditions for the generation and transmission of secondary particles of cosmic rays through the atmosphere change due to the changes in temperature and pressure at different altitudes, resulting in the change of the number of muons that finally reach the earth's surface, which is the meteorological effect of cosmic rays[7,9,13,23]. Therefore, in order to study the cosmic ray muon count on the earth's surface, it is necessary to correct the meteorological effect. The correction formula for the muon count with respect to pressure and temperature can be expressed as

$$I = I' \times f(T) \times g(P), \tag{2}$$

Where $I$ is the corrected muon count, $I'$ is the initial muon count, $f(T)$ is the temperature correction function, and $g(P)$ is the pressure correction function. The temperature and pressure correction functions can be expressed as

$$f(T) = 1 + \alpha_T I_0 \frac{T-T_0}{T_0}, \tag{3}$$

$$g(P) = e^{-\beta \times (P-P_0)}, \tag{4}$$

Where $\alpha_T$ is the temperature correction factor, $I_0$ is the average muon count during the experiment, $T$ is the actual ambient temperature, and $T_0$ is the average ambient temperature during the experiment[24]. $\beta$ is the barometric correction factor, $P$ is the actual barometric pressure, and $P_0$ is the average barometric pressure during the experiment. The daily meteorological data (temperature and pressure) during the experiment were mainly provided by Tongliao Meteorological Bureau, as shown in Fig. 4(a).

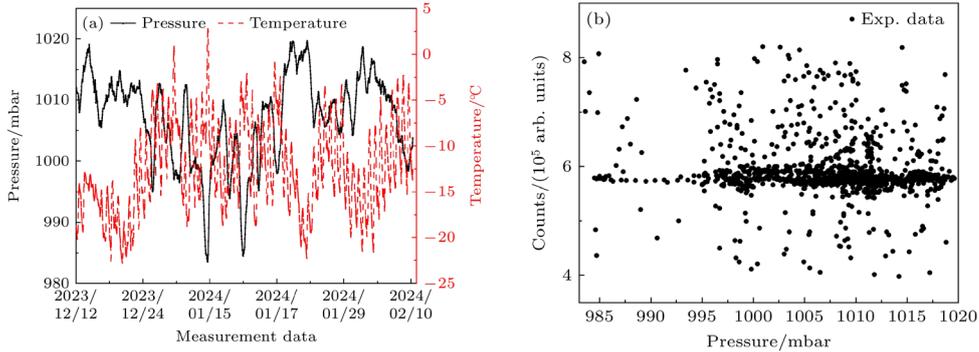

**Figure 4.** (a) Meteorological data information chart during the experiment; (b) count after pressure correction

The atmospheric temperature affects the interaction of primary cosmic rays. With the increase of atmospheric temperature and the decrease of gas density, the probability of primary cosmic ray interaction decreases, which reduces the muon counting rate observed on the surface. According to reference [25], the correlation coefficient between muon count and atmospheric temperature change is $\alpha_T = 0.873$. In this experiment, although the temperature correction function is an important part of the meteorological effect correction, it is found that its impact on the final results is relatively small. This is mainly due to the limited range of temperature fluctuation during the experiment, for example, the difference between the maximum temperature and the minimum temperature during the experiment is relatively small, so that the value of $(T-T_0)/T_0$ does not change much. At the same time, considering the $\alpha_T$ of the temperature correction coefficient and other factors, the change of the muon counting rate caused by the temperature correction is not as obvious as that caused by the pressure correction.

Studies have shown that[26], there is a linear correlation between cosmic ray intensity and ambient pressure, so the linear regression method can be used to calculate the pressure correlation coefficient $\beta$:

$$\beta = r(\sigma_I/\sigma_P), \tag{5}$$

Where $\sigma_I$ and $\sigma_P$ are the standard deviations of intensity and pressure, expressed as $\sigma_I^2 = \frac{1}{N}\sum_{i=1}^{N}(I_i - I_0)^2$, $\sigma_P^2 = \frac{1}{N}\sum_{i=1}^{N}(P_i - P_0)^2$, respectively, and $r$ is the regression coefficient $r = \frac{1}{\sigma_I \sigma_P N}\sum_{i=1}^{N}(I_i - I_0)(P_i - P_0)$. Substituting the Mu counts per hour obtained in the experiment and the pressure parameters at the corresponding time into the (5) formula, and fitting by the least square method, we get $\beta = -0.16$. The relationship between the final corrected barometric pressure and the muon count is shown in Fig. 4(b).

## 3. Results and Discussion

3.1 Cosmic ray muon counting spectrum

The muons measured in this experiment are mainly produced by cosmic galactic rays. The reaction of muons with scintillator produces electrons. Because of the high energy of electrons and the small volume of scintillator, the contribution of high-energy muons in the energy spectrum is small, so the measured counting spectrum is quite different from the energy spectrum in the literature. For a high-energy muon, its specific kinetic energy loss in the plastic scintillator is $2\text{MeV}/\text{cm}$[27], and when it vertically passes through a detector with a thickness of 50 mm, the maximum energy loss is $10\text{MeV}$. The Fig. 5 is the muon counting spectrum measured in this experiment. It can be seen from the figure that there is a Gaussian peak at the $10\text{MeV}$ of the Compton edge, which is consistent with the theoretical calculation of the energy spectrum.

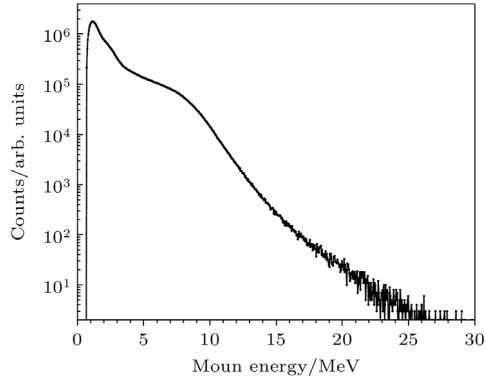

**Figure 5.** Measurement of cosmic ray muon spectra in this experiment

3.2 Solar modulation effect on cosmic ray anisotropy

Because the muons measured in this paper are mainly produced by the reaction of galactic cosmic rays after entering the atmosphere. When galactic cosmic rays enter the solar system and pass through the solar wind region, they will be affected by solar activity and change. These variations include modulations in cosmic ray intensity, direction, composition, and energy spectrum[4]. The Fig. 6 shows the count variation related to the measurement date and time in this experiment. The upper chart shows the count distribution under the time accumulation, and the right chart shows the muon count variation within 24 hours of each day.

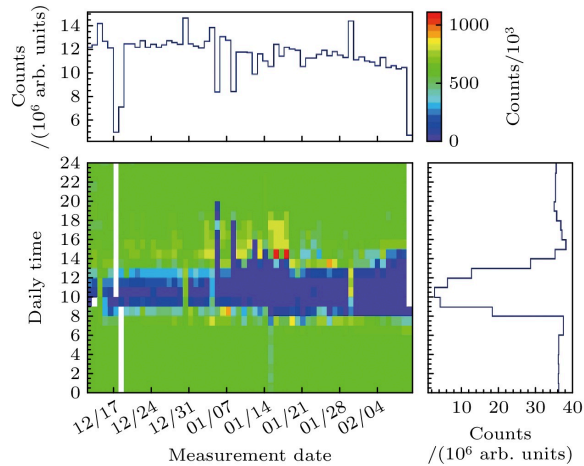

**Figure 6.** The two-dimensional counting spectra of measurement dates and daliy time in this experiment, as well as their projections on the *x* and *y* axes

From the experimental results, it can be seen that from December 2023 to February 2024, the Mu counts at the measurement sites show significant daily periodic variation characteristics. From 8:00 to 13:00 every day, the mule count was significantly lower, while it was relatively higher at other time periods. This counting trend is closely related to the solar modulation effect[6].

As the earth rotates, the relative position of the observation point and the sun changes, resulting in fluctuations in day and night counts. During the daytime from 8:00 to 13:00, the sun is above the observation point, and the sun blocks the galactic cosmic rays from that direction, resulting in a decrease in the muon count during this period. On the contrary, at night, when the observation point is far away from the sun, the cosmic ray muon count increases. The experimental results are in good agreement with the theoretical model, which further reveals the mechanism of the solar modulation effect on the cosmic ray muon distribution at the earth's surface[6].

3.3 Yangbajing data verification

In the research field of cosmic ray muons, Yangbajing Observatory has a high international reputation in cosmic ray observation. Over the years, the observatory has accumulated a large number of high-quality, long-period data, and its authority in the field of cosmic ray research has been widely recognized[6,28,29]. These data provide an important reference for global cosmic ray related research, and can provide a credible benchmark for our experimental results. Therefore, in order to verify the accuracy of the experimental data and the universality of the observation results, the[30] of the observation data of the neutron-muon telescope at Yangbajing Cosmic Ray Observatory in the same time period are selected for comparative analysis (see Fig. 7). The data of Yangbajing observation station also show the characteristics of low counts from 8 o'clock to 12 o'clock every day, and its variation trend is similar to that of the muon detection counts in this experiment, which is related to the "single peak" inverted "U" distribution of solar irradiance in this area. This phenomenon verifies the universality of the anisotropic modulation effect of the sun on cosmic rays in different regions, and to some extent illustrates the effectiveness of plastic scintillator detectors in the observation of cosmic ray muon counting.

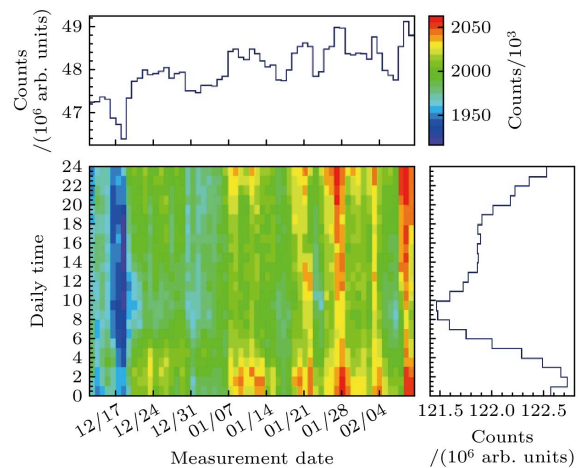

**Figure 7.** The two-dimensional spectrum of the count of the neutron muon telescope at the Yangbajing Cosmic Ray Observatory on the same date, as well as its projection on the *x* and *y* axes

However, due to differences in detector sensitivity and energy range, there are still small differences in count variation at other times throughout the day. The neutron-muon telescope at

Yangbajing station has a high sensitivity to high-energy muons, while the plastic scintillator detector used in this paper has a higher sensitivity to low-energy muons. This difference reflects the difference in the response range of the device and indicates the need for future cross-calibration between multiple detectors in order to detect cosmic ray particles more comprehensively.

The comparison between the cosmic ray muon counts obtained in this experiment and the results of the Yangbajing experiment is shown in Fig. 8. It can be seen from the figure that the normalized cosmic ray muon counts obtained by this experiment and the Yangbajing experiment show certain periodic fluctuations from December 2023 to February 2024. The fluctuation patterns of the two curves are similar in some time periods, for example, from late December to early January, and from mid-January to late January, there are obvious upward and downward trends. Although the overall trend is similar, there are some differences in the specific count values. The data of Yangbajing experiment fluctuates relatively large in most of the time, while the data of this experiment fluctuates relatively gently. This difference may be due to the different geomagnetic cut-off stiffness caused by the different geographical locations of the two experiments. Different geomagnetic cut-off rigidities affect the number and energy distribution of cosmic ray muons arriving at the detector. It may also be the difference of experimental equipment that leads to the difference of counting results. The long strip plastic scintillator detector with double output is used in this experiment, while the neutron-muon telescope detector array is used in Yangbajing experiment.

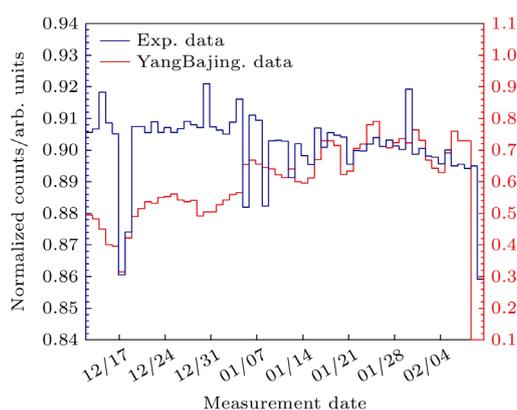

**Figure 8.** Comparison between the normalized cosmic ray muon counts obtained in this experiment and the results of the Yangbajing experiment

# 4. Conclusion

In this study, the counting spectrum of cosmic ray muons and the diurnal variation of the counting rate were measured by a long plastic scintillator detector with double output terminals, which further verified the manifestation of the solar modulation effect in cosmic ray muons.

Through long time observation and double-ended coincidence measurement, we have reduced the noise influence of the single-ended detector and obtained the counting spectrum data of the muon. The experimental results show that the muon counts show significant diurnal periodic fluctuations within 24 hours of a day, especially the lowest counts from 8 o'clock to 13 o'clock every day, which is consistent with the solar shielding effect on galactic cosmic rays, and confirms the influence of solar modulation effect on the observation of surface cosmic ray muons.

In order to verify the reliability of the observation data, the experimental results are compared with the data from Yangbajing Cosmic Ray Observatory, and it is found that the variation trends of the observation data from the two stations are highly consistent, especially in the diurnal variation of the muon count, which further supports the universality of the anisotropic modulation of cosmic rays by the sun, and verifies the reliability of plastic scintillator detectors in the observation of low-energy cosmic ray muons. Based on the advantages of the two-terminal readout in this experiment, it is very important to further optimize the signal processing algorithm of the two-terminal readout. We can learn from the setting and optimization methods of coincidence time window, and explore the use of more advanced digital signal processing technology, such as signal classification and recognition based on machine learning or artificial intelligence algorithms, to improve the efficiency and accuracy of data processing. At the same time, multi-detector joint research, for example, combining different types of detectors (such as plastic scintillator detectors and semiconductor detectors) and using multi-terminal readout technology for cooperative detection, can obtain more comprehensive cosmic ray particle information and provide a more powerful means for in-depth study of the physical properties of cosmic rays.

We would like to thank Professor Zhang Jilong, Institute of High Energy Physics, Chinese Academy of Sciences, for providing key data and useful discussions on the Yangbajing Cosmic Ray Observatory.